\RequirePackage{lineno}

\documentclass[aps,prb,twocolumn]{revtex4}
\usepackage{hyperref}
\usepackage{natbib}
\usepackage{graphicx}
\DeclareGraphicsExtensions{.png,.jpg,.eps,.pdf}

\usepackage{cancel}
\usepackage{xcolor}
\usepackage{amsmath}

\begin{document}

\title{Probing  local moments in nanographenes with electron tunneling spectroscopy}

\author{ R. Ortiz$^{1,2,3}$,  J. Fern\'andez-Rossier$^{3}$\footnote{On leave from Departamento de F\'isica Aplicada, Universidad de Alicante,  Spain} }
\affiliation{(1)Departamento de F\'{\i}sica Aplicada, Universidad de Alicante, 03690,  Sant Vicent del Raspeig, Spain
}
\affiliation{(2)Departamento de Qu\'{\i}mica F\'{\i}sica, Universidad de Alicante, 03690, Sant Vicent del Raspeig, Spain
}
\affiliation{(3)QuantaLab, International Iberian Nanotechnology Laboratory (INL),
Av. Mestre Jos\'e Veiga, 4715-330 Braga, Portugal
}

\date{\today}

\begin{abstract}
The emergence of local moments in graphene zigzag edges,  grain boundaries, vacancies and sp$^3$ defects has been widely studied theoretically.  However, conclusive experimental evidence is  scarce.  Recent progress in on-surface synthesis has made it possible to create nanographenes, such as triangulenes, with local moments in their ground states, and to probe them using scanning tunneling microscope (STM) spectroscopy. Here we review the application of the theory of sequential and cotunneling transport to relate the $dI/dV$ spectra with
the spin properties of nanographenes probed by STM.
This approach permits us to connect the $dI/dV$ with  the many-body  energies and wavefunctions of the graphene nanostructures. We apply this method describing the electronic states of the nanographenes by means of  exact diagonalization of the Hubbard model within a restricted Active Space. This permits us to provide a
proper quantum description of the emergence of local moments in graphene and its interplay with transport.  We  discuss the results of this theory in the case of  diradical nanographenes,  such as triangulene,   rectangular ribbons and the Clar's goblet, that have been recently studied experimentally by means of STM spectroscopy.  
This approach permits us to calculate  both the  $dI/dV$ spectra, that yields excitation energies,  as well as the atomically resolved  conductivity maps, that provide information on the wavefunctions of the  collective spin modes.

\end{abstract}

\maketitle

\section{Introduction}
Ideal graphene, without boundaries and defects,  would be  a diamagnetic zero gap semiconductor with no  unpaired spins. In contrast, both real graphene and graphene nanostructures 
are  expected to  host localized magnetic moments  at edges and defects according to a large amount of theoretical work\citep{nakada1996, fujita1996, ma04,vozmediano05, son2006a, son06b,yazyev07, JFR07, OchoraPRL2008, kumazaki2007,brey07, Palacios08, JFR08, EzawaPRB2007, pereira2008,  boukhvalov08,  yazyev08,castro09, gucclu09,guo09,jung09,lopez09, wang09, Yazyev2010, santos10, soriano11,culchac11,mizukami12, AkhukovPRB2012,  IjasPRB2013,soriano12,golor13a,golor13, Lado2014prl, YazyevNL2014,melle15, Lado15, PhillipsPRB2015,Lado2DShiba, weikprb2016, Lischka2016,Brey2D2017,cao2017,Garcia17, melle17,Ortiz18,Ortiz19,Garcia19}.  Both graphene zigzag edges\citep{nakada1996, fujita1996, JFR07,Lado15,Lischka2016} as well as a large class of $\pi$-conjugated hydrocarbons\citep{morita11,melle15,melle17,Ortiz19} 
 host  zero modes, or singly occupied molecular orbitals, that are prone to host local moments in the $\pi$ orbitals.  Carbon vacancies are expected to have localized spins both in the dangling bonds and the resulting zero mode that arises from the removal of a single $\pi$ orbital from the otherwise ideal graphene\cite{yazyev07,Palacios08,santos10,weikprb2016,Garcia17}.  Analogously,  graphene  functionalized  with sp$^3$ defects, such as atomic hydrogen,  is also predicted to host zero modes with an individual electron,  forming thereby a $S=1/2$ defect \citep{boukhvalov08,santos10}.  Some graphene grain boundaries are also predicted to host zero modes and local moments\citep{lopez09,AkhukovPRB2012, YazyevNL2014, PhillipsPRB2015}, as well as some interfaces between ribbons of different width\citep{cao2017,Ortiz18}. 

All these predictions  are based both on 
density functional theory (DFT) calculations\citep{son2006a,son06b,JFR07,wang09,Yazyev2010,melle15} as well as model Hamiltonian descriptions, at various levels of approximation, going from mean-field approximations\citep{fujita1996,JFR07,JFR08}, spin wave theory\citep{yazyev08,culchac11}, quantum Monte Carlo \citep{golor13a,golor13}, density matrix renormalization group\citep{mizukami12} and exact diagonalizations\citep{gucclu09,Ortiz19}.  Thus, there is a consensus, in the theory front, that graphene local moments arise  in graphene systems where sublattice imbalance is broken\citep{ma04,JFR07,Palacios08,Lieb89}, or in systems where localized states arise close to the Fermi energy, such as grain boundaries\citep{lopez09,AkhukovPRB2012, YazyevNL2014, PhillipsPRB2015}.   These local moments are predicted  to have  very interesting properties:  a very elegant interplay between sublattice and spin polarization\citep{brey07,JFR07,JFR08,soriano12},  prone to electrical control\citep{son06b,castro09,Garcia19} and electrically driven spin resonance\citep{Ortiz18}, exotic Yu-Shiba-Russinov states when proximitized by a superconductor\citep{Lado2DShiba}, domain walls with fractional charge\citep{Brey2D2017},   and even potential for  quantum computing\citep{guo09}.
 
 The situation in the experimental front is  less advanced
  due to several  reasons. First, most of the experiments rely on ensamble magnetism measurements of nanographenes (NG)\citep{inoue01},   defective graphene \citep{nair13,makarova15}, and electrically detected spin resonance\citep{lyon17}, and these   measurements can be prone to artefacts arising from the presence of extrinsic magnetic impurities\citep{sepioni12}.    Second,  bottom-up synthesis of open shell nanographenes is in general a low yield process, on account of the strong chemical reactivity of radical species\citep{morita11}.
  Third, the fabrication of structures with  well defined zigzag edges, using top-down techniques  was very challenging\citep{ozyilmaz07}, on account of the  lack of sufficient precision provided by chemical etching.  A significant step forward in this direction was made possible by creating ribbons  unzipping  carbon nanotubes, that  made it possible to study graphene ribbons with atomically defined edges using scanning tunneling microscopy (STM) \citep{tao11}. 
 
Progress in bottom-up on-surface synthesis\citep{cai2010} in ultra-high vacuum  has  made it possible
 to synthetize graphene nanostructures such as graphene ribbons with zigzag edges\citep{KimoucheNat2015,wang2016,Ruffieux16}, expected to have edge magnetism with antiferromagnetic correlations between opposite edges\citep{nakada1996,son06b,JFR08,jung09}, triangulenes with zigzag edges \citep{pavlivcek2017,su19,mishra19}, expected\citep{JFR07} to have ferromagnetic ground state, and nanoribbon heterojunctions that host localized zero modes \citep{RizzoNat2018, GroningNat2018,CrommieACSnano2018} that are expected to host local moments \citep{cao2017,Ortiz18}.  Atomic manipulation of individual atomic hydrogen chemisorbed on graphene, and the resulting changes in $dI/dV$ observed with STM, have been reported \citep{Sciencejuanjo} and interpreted in terms of the emergence of a localized spin. 
 
 In most of these experiments \citep{wang2016,Ruffieux16,pavlivcek2017,su19,mishra19, RizzoNat2018,GroningNat2018,CrommieACSnano2018,Sciencejuanjo}, 
 STM $dI/dV$ provides spectra with broad peaks, without positive-negative bias  symmetry,  that are interpreted in terms of the tunneling to the HOMO-LUMO levels of the structures.  In some cases, comparison with DFT and GW calculations yields a fairly good agreement  \citep{tao11,Sciencejuanjo,wang2016}. The fact that  these calculations predict the existence of local magnetic moments provides indirect evidence for the emergence of magnetism in these systems.  
 
A much more direct evidence of local moment emergence is the observation of {\em steps} in the $dI/dV$  spectra,   observed at $|e|V_{bias}=\pm \Delta$, where $\Delta$ is the excitation energy. If $\Delta$  depends on the applied magnetic field, this automatically implies that it corresponds to a spin excitation,  observed in single magnetic atoms on surfaces \citep{Heinrich04,Hirji07,khajetoorians11}, in nano-engineered adatom chains\citep{Hirji06,spinelli14}, or single magnetic molecules \citep{Chen08,Tsukahara09,burgess15}. In some instances,  variations of the intensity of a given  inelastic step across different atoms in a given structure are observed, providing information of the  wavefunction of the spin excitation\citep{spinelli14,toskovic16} .  These symmetric inelastic steps can be understood in terms of inelastic cotunneling theory\citep{DelgadoPRB2011}, and they probe the energy difference between two many-body states with total spin $S_1$ and $S_2$, with $|S_1-S_2|=0,\pm 1$ and $|S_{z1}-S_{z2}|=0,\pm 1$. 
Supplemented with theory\citep{JFR09,gauyacq12}, these experiments  permit to infer the spin Hamiltonian of the system. 

STM inelastic spectroscopy  showing features compatible with cotunneling steps have been reported in 
 nanographenes  of various shapes with zigzag edges,  such as fused ribbons  \citep{NachoNat,li2019} and the Clar's goblet\citep{ManuelNewsandviews,mishra19b}.
 Further evidence of the emergence of local moments in these structures arises from 
 the controlled addition of either atomic hydrogen\citep{NachoNat,li2019} or pentagonal defects \citep{mishra2020} that leads to the appearance of the Kondo peak.
    In the case of fused ribbons\citep{NachoNat} and a triangulene fused to an acene\citep{li2019}, the appearance of the Kondo effect is accompanied by  the  disappearance of the inelastic steps,  compatible with a transition of the ground state from $S=0$ to $S=1/2$ in the fused ribbon and $S=1$ to $S=1/2$ in the triangulene.  
Intriguingly, cotunneling  steps are conspicuously missing in the spectra of triangulenes\citep{pavlivcek2017,su19,mishra19} and graphene ribbons with short zigzag edges\citep{wang2016,Ruffieux16}.  Possible reasons for this negative observation are discussed below.
\begin{figure}[t]
 \centering
  \includegraphics[width=0.47\textwidth]{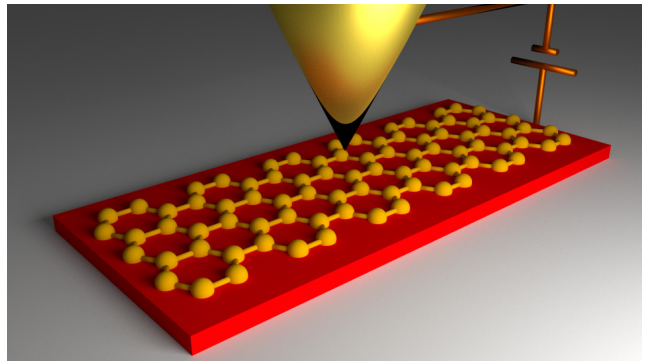}
\caption{Scheme of the system of interest: an STM tip on top of a nanographene deposited on a  decoupling layer (red) on top of a conducting substrate (grey).}
\label{fig1}
\end{figure}

The goal of this paper is to review the theoretical background for
 all this work and to elucidate to what extent   tunneling spectroscopy can
probe local moments in many graphene nanostructures.    For that matter, we compute  the contributions to $dI/dV$ of both sequential tunneling and inelastic cotunneling and we relate them to the nature of the many-body states of nanographenes deposited on surfaces.  The multielectronic states of nanographenes are obtained by exact diagonalization of the Hubbard model in a restricted configuration Hilbert space and the coupling to both tip and substrate is treated to lowest order in perturbation theory.  

The rest of this paper is organized as follows. In section II we review the relevant energy scales for  sequential tunneling and cotunneling and we present an extended Hubbard model to describe nanographenes.  In section III we review the formalism for sequential tunneling and cotunneling transport. We illustrate sequential transport theory with calculations for the case of a rectangular nanographene with edge modes.  In section IV we apply cotunneling theory to the case of three different diradical nanographene structures,  for which there are recent experiments:  a rectangular nanographene\citep{wang2016},  triangulene\citep{pavlivcek2017,melle17} and the Clar's goblet\citep{ManuelNewsandviews,mishra19b}.
In section V we discuss a number of open questions and aspects in which further theory work is needed. In section VI we wrap up the main conclusions.

\section{Theory}
The main goal of this work is to describe, theoretically, electronic transport between an STM tip and a nanographene deposited on a conducting substrate (see figure \ref{fig1}). In the following we {\em assume} that  electron tunneling events between the nanographene and both tip and substrate are weak. Whereas this is a realistic assumption in the case of the graphene-tip coupling, tunneling to the substrate may not be weak.  The weak coupling assumption is expected to work better in cases when graphene is separated from the conducting substrate by a decoupling insulating layer. This is the case of some of the experiments\citep{wang2016,pavlivcek2017}, but in most cases nanographenes are deposited directly on a metallic substrate, in which case the weak coupling theory may not be applicable, but can still be used as starting point\citep{ervasti17}.

\subsection{Sequential transport energy scales}
Within the weak coupling hypothesis, current flows via tunneling events of two types: sequential and cotunneling.  Sequential processes entail classical charge fluctuations of the nanographene: an electron tunnels from tip to the nanographene, which is charged until another tunneling event takes an electron from the nanographene to the substrate.
Since the nanographene is directly deposited on the metal  surface, the efficiency of the Sequential Transport (ST) processes is governed by the tip-graphene  tunnel process, that acts as a bottleneck.    
Energy conservation demands that:
\begin{equation}
\mu_T+ E_G(N)=E_G(N+1)
\label{eq1}
\end{equation}
where the left hand of  this equation is the energy of the system with a quasiparticle at the tip Fermi energy ($\mu_T$) and the nanographene with $N$ electrons, and the right hand side is the energy when the quasiparticle has entered the nanographene.

As the bias voltage ($V_{bias}$) is ramped, it will shift the chemical potential of the electrodes ($\mu_{\eta}$), where $\eta\in S,T$. This shift can be assymmetric\citep{LossPRB2004}, and here we will assume that the bias will just affect $\mu_T$ in accordance
to the much larger capacitance of the surface. We thus assume:
\begin{equation}
\mu_T = \mu_S+|e|V_{bias}
\label{eq2}
\end{equation}
 where $\mu_{S,T}$ are the chemical potentials of substrate and tip, including the contact formation corrections when tip and sample are made of different materials.  Since tip and substrate are conducting, we can identify chemical potential and work function.  Combining equations (\ref{eq1}) and (\ref{eq2}) we conclude that sequential transport is allowed  for the (positive) bias voltage $V_{+1}$: 
 \begin{equation}
 E_+=|e|V_{+1}= E_G(N+1)-E_G(N)-\mu_S
\label{eq3}
\end{equation}

Signs are chosen so that for positive $V_{bias}$, electrons move from tip to sample. For negative bias voltage, sequential transport events are controlled by processes in which an electron in an occupied level of the nanographene  tunnels towards the tip.  The minimal (negative) voltage $V_{-1}$ at which this occurs is given by equation
\begin{equation}
E_G(N)=\mu_T+E_G(N-1)=\mu_S-|e V_{-1}|+E_G(N-1)
\end{equation}
which leads to the condition
 \begin{equation}
 E_-=|e V_{-1}|=  E_G(N-1)-E_G(N)+\mu_S
\label{eq5}
\end{equation}

In this work, we compute these quantities using an exact diagonalization of the Hubbard model for nanographenes.
We now briefly discuss how to estimate the addition voltages in an independent particle picture,   such as non-interacting electron approximation, Hartree Fock, and density functional based calculations. Under this framework, we can write up:
\begin{eqnarray}
E_G(N+1)-E_G(N)&=&\epsilon_{\rm LUMO} \nonumber \\
E_G(N)-E_G(N-1)&=&\epsilon_{\rm HOMO}
\label{MO}
\end{eqnarray}
where LUMO and HOMO stand  for lowest unoccupied and highest occupied molecular orbitals. Thus, if equations (\ref{MO}) hold,  we would write: 
\begin{eqnarray}
 |e|V_{+1}= \epsilon_{\rm LUMO}-\mu_S\nonumber\\
  |e V_{-1}|= \mu_S -\epsilon_{\rm HOMO}
\label{HL}
\end{eqnarray}
Within this picture, every time the positive (negative)  bias aligns with an empty (occupied) state,  sequential tunneling processes are possible.  Thus, we can expect peaks in the $dI/dV$  when $V_{bias}$ goes across these resonances. 

 It is important to notice that 
unless the addition and substraction energies ($E_\pm$) are the same,  the voltages $V_{+1}$ and $V_{-1}$ will be different in magnitude   
and, hence,  sequential tunneling $dI/dV$ peaks are not symmetrically located around $V_{bias}=0$.
This is clearly the case of the experiments of nanographenes placed on metals \citep{wang2016, ErvastiJESRP2017, pavlivcek2017}, see Table (\ref{table1}).  

In the non-interacting picture,  HOMO and LUMO levels are symmetrically placed around the work function  of graphene, $\epsilon_{LUMO}=\epsilon_C+ \frac{\delta}{2}$,  
$\epsilon_{HOMO}=\epsilon_C- \frac{\delta}{2}$, where $\delta$ is the single-particle gap, and we can write:
\begin{eqnarray}
 |e|V_{+1}= \epsilon_C +\frac{\delta}{2} -\mu_S\nonumber\\
  |e V_{-1}|= \mu_S -(\epsilon_{C}-\frac{\delta}{2})
\label{HL2}
\end{eqnarray}
So, the electron-hole symmetry condition, $|e|V_{+1}=|e V_{-1}|$ would happen only if $\epsilon_C=\mu_S$.  This situation might occur for nanographenes deposited on graphene, as long as the interaction effects are negligible.  

In the case where single-particle theory predicts a singly occupied zero mode, whose wavefunction is denoted by $\phi_0$,  the Coulomb overhead of adding a second electron in that orbital is given in the Hubbard approximation by\citep{IjasPRB2013,Ortiz18,Ortiz19} 
\begin{equation}
\tilde{U}=U\sum_i |\phi_0(i)|^4
\end{equation}
which is a metric of the orbital delocalization.
This  applies  for the case of monohydrogenated graphene\citep{Sciencejuanjo}, very long rectangular ribbons with  zigzag  edges on the short side\citep{IjasPRB2013,wang2016},   and triangulenes\citep{pavlivcek2017,su19,mishra19}, where transport occurs via  zero modes. In this case, we have:

 \begin{eqnarray}
 |e|V_{+1}&\simeq& \tilde{U} -(\mu_S-\epsilon_C )\nonumber\\
  |e V_{-1}|&\simeq&\mu_S -\epsilon_{C}
\label{HL3}
\end{eqnarray}

Thus, in the case of transport through zero modes, the sum of the absolute values of the addition and substraction peaks is a metric of the Coulomb overhead associated to the double occupancy of the zero mode,  and thereby its spacial extension. The  compilation of experimental results shown in  Table\ref{table1} suggests that  $\tilde{U}$  is minimal for the zero mode associated to hydrogenation of 2D graphene, expected from theory\citep{Garcia17}, and maximal for the triangulene with 22 carbon atoms.

 \begin{table}[htp]
 \caption{Experimental position of Sequential Transport peaks for several nanographenes}
 \begin{center}
    \begin{tabular}{| l | l | l | }
    \hline
    Structure & $E_+(eV)$ & $E_-(eV)$\\ \hline
    [6]ribbon (NaCl/Au[111])\citep{wang2016} & 1.3 & 0.5  \\ \hline
    [3]triangulene (Xe[111])\citep{pavlivcek2017} & 1.85 & 1.4  \\\hline
    [4]triangulene (Au[111])\citep{mishra19} & 1.15 & 0.4 \\ \hline
    [5]triangulene (Au[111])\citep{su19} & 1.07 & 0.62 \\ \hline
    H + graphene (SiC)\citep{Sciencejuanjo} & 0.014 & 0.007 \\  \hline
    Clar's goblet (Au[111])\citep{mishra19b} & 1.0 & 0.3 \\
    \hline
    \end{tabular}
\end{center}
\label{table1}
 \end{table}

\subsection{Cotunneling energy scales}

In addition to sequential tunneling events, transport can also occur via cotunneling\citep{de01}: an electron enters(leaves) the graphene nanoisland, initially in the state $E_i(N)$, turning it towards  an excited state $E(N\pm1)$ during a Heisenberg time, and coherently in a second tunneling event a second electron steps out(in), in the other electrode, and the island stays in the state $E_f(N)$.  This process can be both elastic or  inelastic, depending on whether $\Delta=E_f(N)-E_i(N)=0$ or else. Total energy cannot change between initial and final state. Therefore,   inelastic processes are possible when bias voltage matches the inelastic energy:
\begin{equation}
 |eV_{cot}|\geq \Delta 
 \label{cotuncondition}
\end{equation}

Thus, as  $|e|V_{bias}$ is  increased, new inelastic cotunneling channels open, increasing the conductance in a step-wise manner. This is the principle of cotunneling spectroscopy.  Cotunneling contribution to conductance is in general much smaller than sequential processes, on account of its non-resonant nature. Therefore,  it is better observed when the inelastic steps are away of the ST peaks. Therefore, a condition for the cotunneling steps to be  observable is that their energy is much smaller than the resonant peaks $V_{\pm 1}$.  As we discuss below, this condition is not always satisfied.

Importantly,  cotunneling events include both spin-flip and spin conserving  processes\citep{DelgadoPRB2011}.  Spin conservation entails $\Delta S=\pm1$ or $\Delta S=0$.  If $\Delta S_z=\pm 1$,  $\Delta$ should depend on a magnetic field applied along the $z$ axis. Therefore,  the magnetic field dependence of the inelastic steps provides an unambiguous proof of the spinful nature of at least one of the two many-body states implied in the excitation.

\subsection{Hamiltonian}
\begin{figure}[h!]
 \centering
  \includegraphics[width=0.48\textwidth]{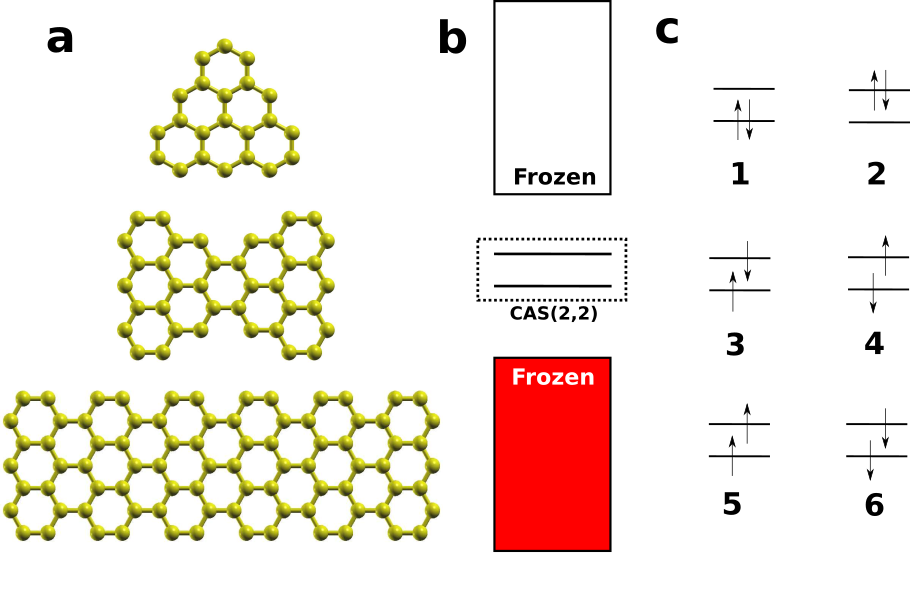}
\caption{a) Atomic structure of the three systems  studied in this work. From top to bottom: triangulene, Clar's goblet and a rectangular ribbon.  Hydrogen atoms that passivate 
edge carbons are not shown for clarity.
b) Scheme of  CAS approximation.   The red and white boxes represent the non-active spaces, with molecular orbitals that are doubly occupied and empty, respectively. The Active Space is marked by states within the box.  c) The 6 possible configurations in the CAS(2,2) approximation.}
\label{fig2}
\end{figure}

We now introduce an extended Hubbard model Hamiltonian. The systems of interest, shown in figure (\ref{fig2}a),  are 3 types of nanographenes that are either diradical, like the triangulene and the Clar's Goblet, or have hybridized zero modes, like the rectangular ribbon.  The  single-particle part of the Hamiltonian is the standard one orbital tight-binding model  with first neighbour hopping $t$ for the $\pi$ orbitals.  Edge atoms are assumed to be passivated with hydrogen.  Electron-electron interaction is treated with two terms. First, we add  an on-site Hubbard repulsion $U$:
\begin{equation}
{\cal H}_{\rm Hub} = t\sum_{\langle i,j \rangle,\sigma}( c^{\dagger}_{i,\sigma} c_{j,\sigma} + h.c.) +
U  \sum_{i}  n_{i\uparrow } n_{i\downarrow }
\label{hub}
\end{equation}

We denote the ground state energy   of the manifold with $N$ electrons  for the Hubbard model  (\ref{hub})
by $E_G(N)$.
The Hubbard model ignores completely the long range part of the Coulomb interaction. As a result, its addition  energies ($E_G(N\pm1)-E_G(N)\mp \mu_S$) 
are not correctly captured by the model. Second, in order to address this shortcoming, we adopt an heuristic solution and add an extra term in the 
Hamiltonian (equation \ref{lambdaH}) that yields the correct energies:

\begin{equation}
 {\cal H}_\lambda = \frac{\lambda_++\lambda_-}{2} \left(\hat{N}-N_s\right)^2+
 \frac{\lambda_+-\lambda_-}{2}\left(\hat{N}-N_s\right)
\label{lambdaH}
\end{equation}

where  $\hat{N}=\sum_{\sigma,i} c^{\dagger}_{i,\sigma} c_{i,\sigma} $ is   the number operator, $\lambda_{\pm}$   are two phenomenological parameters chosen to make sure that the addition energies match those seen in the experiment, and $N_s$ is the number of carbon sites in the nanographene.

 In addition, we have the Zeeman coupling and the onsite energy for carbon atoms:
\begin{equation}
 {\cal H}_C+ {\cal H}_{Zee}= \epsilon_C\sum_{i\sigma}c^{\dagger}_{i\sigma}c_{i\sigma}
  + \frac{1}{2}g\mu_B\vec{B}\cdot\vec{\sigma}
\end{equation}
where $\vec{B} = (B_x,B_y,B_z)$ is an external magnetic field, $g=2$, $\mu_B$ is the Bohr magneton and $\vec{\sigma}$ is the Pauli matrices vector.

The STM tip is known to pull the atoms out\citep{WongACSnano2009, KochNatNano2012, XuPRB2012, XuPRB2012bis, MorgensternNL2017}. As a result the  hopping energies  of the atoms underneath the tip are reduced because of the misalignment of the $\pi$ orbitals. We assume that this deformation only affects to the one atom right under the STM tip, labeled with the index $0$:
\begin{equation}
{\cal H}_{\rm def}= \delta t  \sum_{\langle0, j \rangle,\sigma}( c^{\dagger}_{0,\sigma} c_{j,\sigma} + h.c.)
\label{def}
\end{equation}

In the following, unless otherwise stated, we take $\delta t=-0.3 t$. This perturbation has a minor impact in both the single-particle and many-body spectra, but, as we discussed below, it opens the otherwise closed cotunneling channels for triangulenes.

Thus,  the Hamiltonian  for the nanographenes  is the sum of the Hubbard model, with the on-site and Zeeman terms, plus the charging energy and the tip deformation corrections:
 \begin{equation}
 {\cal H}_{NG} = {\cal H}_{\rm Hub}+ {\cal H}_C + {\cal H}_{Zee}+{\cal H}_\lambda+{\cal H}_{def}
 \label{HAMIL}
 \end{equation}
In the following we label the eigenstates of ${\cal H}_{NG}$ as $|n\rangle$ for the manifold with $N_s$  $\pi$ electrons and as $|m_\pm\rangle$ for the eigenstates with $N_s\pm 1$.

\subsection{Complete Active Space approximation}
As we discuss below, the transport calculations take as an input the multi-electronic eigenstates of (\ref{HAMIL}).
We solve the Hubbard model with the Configuration Interaction (CI)  method in the Complete Active Space (CAS) approximation. 

In the CAS method, we break down the single-particle spectra, obtained from the diagonalization of the single-particle part of the Hamiltonian (\ref{HAMIL})  in 3 sectors: low energy sector, Active Space  and high energy sector (figure \ref{fig2}b). 
We build a basis of many-body  Fock  states with well defined (0,1) occupation of the single-particle basis.  In all the Fock states in the basis, the occupation of the low (high) energy sectors is 1(0), and they differ in the occupation of the Active Space.  A  CAS basis is defined by the number of electrons $N_e$ and the number of molecular orbitals $N_o$ (without accounting for spin degeneracy) in the Active Space.  For the charge neutral manifold $|n\rangle$  we use $N_e=N_o=2$ or $N_e=N_o=4$, which leads to a CAS basis of dimension $\left( \begin{smallmatrix} 2N_o\\ N_e \end{smallmatrix} \right) = \left( \begin{smallmatrix} 4\\ 2 \end{smallmatrix} \right) =  6$ (see figure \ref{fig2}c) and $\left( \begin{smallmatrix} 2N_o\\ N_e \end{smallmatrix} \right) = \left( \begin{smallmatrix} 8 \\ 4 \end{smallmatrix} \right) =70$, respectively. For $N_s\pm 1$ manifolds we add or remove one electron.

Once the CAS basis is defined, we represent the many-body Hamiltonian  in that basis and diagonalize. This gives us an approximate description of the eigenstates of ${\cal H}_{NG}$. Importantly, the quantum states obtained in this  approach provide a full quantum description of magnetism,  preserve the spin rotational invariance of the Hamiltonian, and capture quantum spin fluctuations, unlike the broken symmetry picture of  the mean-field approximation. For instance, the $S=0$ singlets  are entangled states that combine antiferromagnetically correlated electrons that are linear superposition of $\uparrow$ and $\downarrow$ states.

\section{Transport}
In this section we review the theory of both sequential and contunneling transport. The starting point is the definition of a Hamiltonian that includes both the nanographene Hamiltonian, presented in the previous section, the tip and substrate Hamiltonians, and their coupling (see for instance 
\citep{LossPRB2004,SobczykPRB2012,gaudenzi17}): 
\begin{equation}
 {\cal H} = {\cal H}_T + {\cal H}_S  + {\cal H}_{NG}  + {\cal V} 
 \label{trans}
\end{equation}
The first two terms describe the electrons in tip and substrate. We treat them in the independent electron approximation and we label their fermions with the operators $(f_{\alpha}^{\dagger},f_{\alpha})$, where  $\alpha \equiv \{ \eta, k, \sigma \}$,  so $\eta=S,T$ labels the electrodes, $k$ the momentum and $\sigma$ the spin of the quasiparticles.
 The third term  {${\cal H}_{NG}$, given by  equation (\ref{HAMIL}), describes the nanographene.  
 
 The last term, ${\cal V}$,  describes the tunneling of electrons from the tip and the substrate to the nanographene:
 \begin{eqnarray}
  {\cal V}= \sum_{\alpha,i}(V_{\alpha}(i) f^{\dagger}_{\alpha}c_{i\sigma}+h.c.)={\cal V}_S+ {\cal V}_T
 \end{eqnarray}
where $V_{\eta,k,\sigma}(i)$ stands for the hopping matrix element that connects a state in electrode $\eta$, with spin $\sigma$ and single-particle quantum number $k$, and the $\pi$ atomic orbital of the carbon site $i$ in the nanographene. Quasiparticle spin is conserved in the tunneling processes. In the case of tip-nanographene tunneling, the matrix element depends strongly on $i$. We tipically assume that only one carbon atom is coupled to the tip. In contrast, unless otherwise stated, we assume that all carbon atoms are equally coupled to the substrate states, so that $V_{S,k,\sigma}$ does not depend on the site index $i$.

\subsection{Sequential tunneling}
The calculation of current in the sequential tunneling approximation treats ${\cal V}$ in perturbation theory.  The energy conserving tunneling rates that connect the neutral states of the nanographene ($|n\rangle$) with the charged states ($|m_\pm\rangle$) are given by the
Fermi's golden rule\citep{LossPRB2004,EzawaPRB2008,gaudenzi17}, that emits/receives the tunneling quasiparticle:

\begin{equation}
 \Gamma^\eta _{n\rightarrow m_+} = \frac{2\pi\rho_{\eta\sigma}}{\hbar}n_F(\xi_\eta)\sum_{i,i',\sigma}V_{\eta\sigma}(i)V^*_{\eta\sigma}(i')
\gamma^{m_+}_{nn}(ii'\sigma\sigma)
\label{gamma1}
\end{equation}
\begin{equation}
 \Gamma^\eta _{m_+\rightarrow n} = \frac{2\pi\rho_{\eta\sigma}}{\hbar}n_F(-\xi_\eta)\sum_{i,i',\sigma}V_{\eta\sigma}(i)V^*_{\eta\sigma}(i')
  \gamma^{m_+}_{nn}(ii'\sigma\sigma)
\label{gamma2}
\end{equation}
\begin{equation}
 \Gamma^\eta _{n\rightarrow m_-} = \frac{2\pi\rho_{\eta\sigma}}{\hbar}n_F(\xi_\eta)\sum_{i,i',\sigma}V_{\eta\sigma}(i)V^*_{\eta\sigma}(i')
  \gamma^{m_-}_{nn}(ii'\sigma\sigma)
\label{gamma3}
\end{equation}
\begin{equation}
 \Gamma^\eta _{m_-\rightarrow n} = \frac{2\pi\rho_{\eta\sigma}}{\hbar}n_F(-\xi_\eta)\sum_{i,i',\sigma}V_{\eta\sigma}(i)V^*_{\eta\sigma}(i')
  \gamma^{m_-}_{nn}(ii'\sigma\sigma)
\label{gamma4}
\end{equation}

where $\xi_\eta = E_{m_\pm} -E_n \mp \mu_\eta$, $\rho_{\eta\sigma}$ is the electrode density of states, $n_F$ is just the Fermi-Dirac distribution function, and $V_{\eta\sigma}(i)$ are the electrode-NG matrix elements, neglecting their dependence on the quasiparticle label $k$.  The $\gamma$ matrices encode the  height of the peaks in the nanographene {\em fermion spectral function}:
\begin{eqnarray}
 \gamma^{m_+}_{nn'}(ii'\sigma\sigma')= \langle n|c_{i\sigma}|m_+\rangle\langle m_+|c^\dagger_{i'\sigma'}|n'\rangle\nonumber\\
 \gamma^{m_-}_{nn'}(ii'\sigma\sigma')= \langle n|c^\dagger_{i\sigma}|m_-\rangle\langle m_-|c_{i'\sigma'}|n'\rangle
 \label{gamma}
\end{eqnarray}

The scattering rates (\ref{gamma1},\ref{gamma2},\ref{gamma3},\ref{gamma4}) hence define transitions between states with different charge that will be activated by bias when it matches the
addition or substraction energies. 
The dynamics of the occupation of the NG many-body states is governed by a master equation:
\begin{equation}
 \frac{dP_l}{dt} = -(\sum_{l\neq l'}\Gamma_{l\rightarrow l'})P_l + \sum_{l'\neq l}\Gamma_{l'\rightarrow l}P_{l'}
 \label{master}
\end{equation}
where $l$ labels the states $n,m_{\pm}$.

Here we are interested in the case when the stationary state is reached, i.e. when $\frac{dP}{dt} = 0$ and the intensity is the same at both
electrodes ($I_T=I_S$). This leads to the expressions:
%

\begin{eqnarray}
 I_{T}= e\left(
 \sum_{m_{+} }P_{m_+}\sum_{n}\Gamma^T_{m_+\rightarrow n}-\sum_n P_{n}
  \sum_{m_{+}}\Gamma^T_{n\rightarrow m_+}\right)
 \label{intensityseq}
\end{eqnarray} 

or

\begin{eqnarray}
 I_{T}= e\left(\sum_n P_{n}
  \sum_{m_{-}}\Gamma^T_{n\rightarrow m_-}
 -\sum_{m_{-} }P_{m_-}\sum_{n}\Gamma^T_{m_-\rightarrow n}\right)
 \label{intensityseq}
\end{eqnarray}

so for positive (negative) bias the charge fluctuations occur via transitions betwen the $N=N_s$ manifold with the $m_+$ ($m_-$) states.

\begin{figure}[h!]
 \centering
  \includegraphics[width=0.50\textwidth]{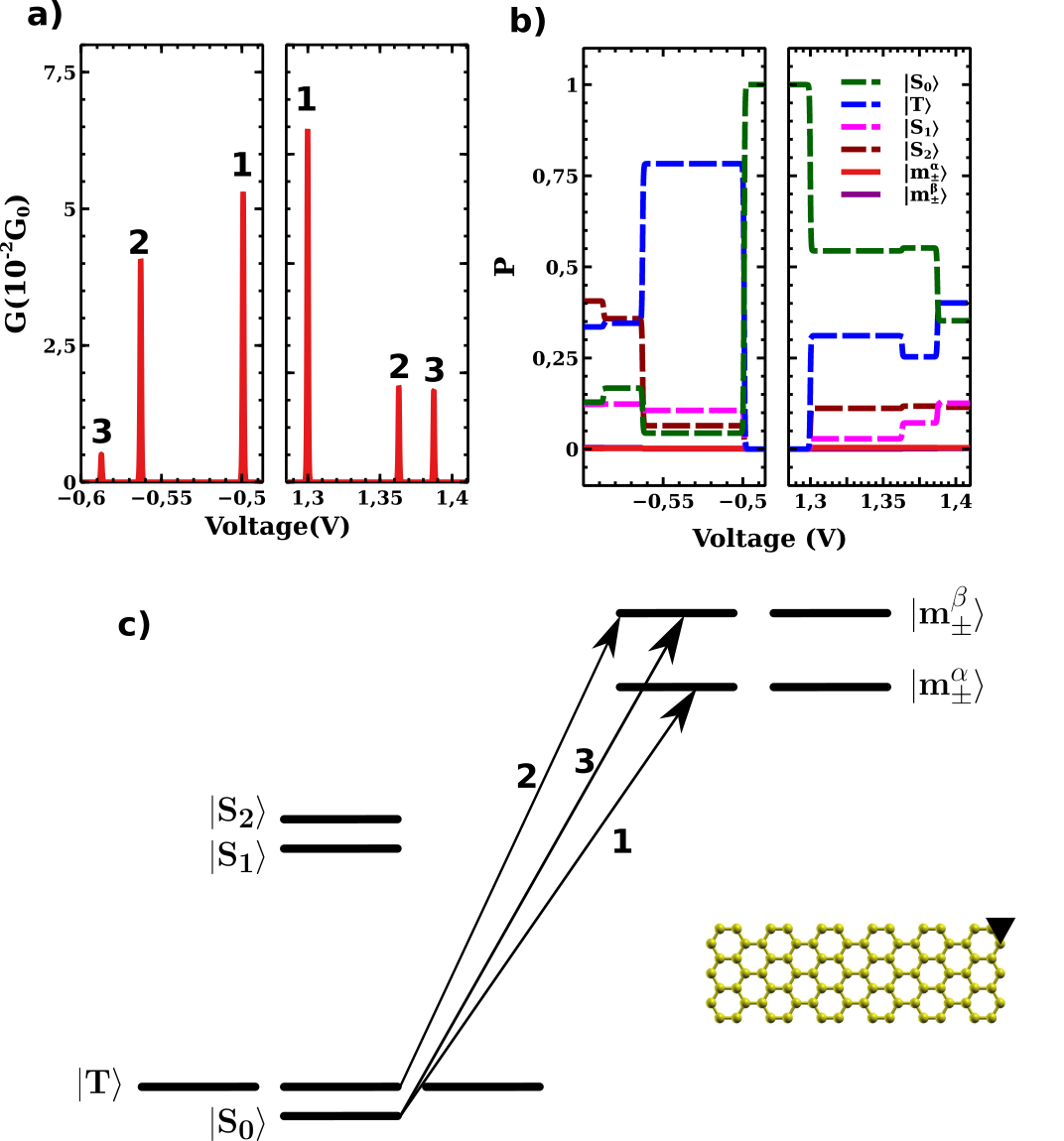}
\caption{a) Conductance curves for sequential transport for the rectangular nanographene
shown in the inset of panel (c) and the tip coupled to
an edge atom, marked with the triangle symbol. $G_0$ is the quantum of conductance.
b) Population of the many-body states of neutral and charged manifolds as a function of bias, obtained from 
the steady state solution of  equation (\ref{master}).
c) Scheme of the CAS(2,2) many-body energy levels  for the $N=N_s$ (neutral) manifold and
the $N=N_s\pm1$ manifolds. The numbered arrows mark the  transitions that result in the conductance peaks in panel (a). 
$t=-2.7eV$, $\delta t=-0.3 t$, $U = |t|$, $\epsilon_C=-5.7eV$, CAS(2,2), $T=3K$, $\rho_{\eta\sigma}=10/t$, $V_S = t/100$ and $V_T = V_S/10$.}
\label{fig3}
\end{figure}

We now  apply the theory 
to the case of a rectangular graphene flake with well defined edges, that has been studied experimentally\citep{wang2016}.
In figure (\ref{fig3}a,b) we show the calculated $dI/dV$ curve  and the population of the many-body states as a function
of bias. For a given polarity, we obtain three peaks. 
The lowest energy peaks, labeled with $\bf 1$, correspond to
the addition and removal energies  given by  equations (\ref{eq3}) and (\ref{eq5}).  These peaks involve transitions between two ground states of manifolds with different charge. 
The  values of $\lambda_{\pm}$ are chosen so that the first addition and substraction energies ($E_\pm$) 
match the experimental observation\citep{wang2016} (see Table\ref{table1}). Expectedly, the peak positions are not electron-hole symmetric. 

The higher energy peaks (labeled with ${\bf 2}$, ${\bf 3}$) correspond to ground to excited transitions (${\bf 3}$) and 
excited to excited transitions ($\bf 2$):
\begin{equation}
\xi_S = |eV_{m_\pm,n}|= E_{m_\pm}-E_n\mp\mu_S
\end{equation}

In figure (\ref{fig3}b) we show the occupation of the  states from manifolds with $N=N_s$, and $N=N_s\pm 1$ as a function of bias, obtained by solving the master equation (\ref{master}).  It is apparent  that the peaks in the $dI/dV$ occur at the same bias for which the charge of the nanographene fluctuates.  Thus,  transport is enabled by a combination via {\em classical} charge fluctuations and tunneling events between nanographene and the electrodes.

In the framework of sequential transport theory, the width of the peaks is controlled by  temperature.  However, the broadening observed in experiments is dramatically larger.
Other than thermal smearing of the electrode quasiparticles, the ingredients that contribute to broadening, missing in sequential transport theory, are two.
  First,  since  coupling to the electrodes is treated at the lowest order, the spectral function of the nanographene is made of infinitely narrow peaks.  Thus, coupling to the substrate induces quasiparticle broadening\citep{ervasti17,jacob18}, not captured in the sequential tunneling approach.
Second,  and definitely relevant in the case where a polar decoupling layer such as NaCl separates graphene from the substrate\citep{ReppPRL2005,wang2016}, the decoupling layer vibrations are known to broaden the resonant tunneling peaks \citep{ReppPRL2005b}}. In any event, this problem deserves further attention.

\subsection{Cotunneling formalism}
We now briefly review the cotunneling formalism. We follow previous work by one of us
\citep{DelgadoPRB2011}.  The first step in the method entails the derivation of a new tunneling Hamiltonian where the charged states of the nanographene with $N\neq N_s$, i.e. the manifolds $|m_{\pm}\rangle$, are integrated out. This leads to an {\em effective} tunneling Hamiltonian where quasiparticles tunnel directly from tip to substrate, inducing transitions between the many-body states of the nanographene in the $N=N_s$ manifold: 
\begin{equation}
 {\cal H}_{cotun}= \sum_{\alpha\alpha'}[ \hat{{\cal O}}^{(+)}_{\alpha\alpha'} - \hat{{\cal O}}^{(-)}_{\alpha'\alpha}]f^{\dagger}_{\alpha}f_{\alpha'}
 \label{cotun}
\end{equation}
where the operator
\begin{equation}
 \hat{{\cal O}}^{\pm}_{\alpha\alpha'} \equiv \sum_{nn'} \langle n|\hat{{\cal O}}^{(\pm)}_{\alpha\alpha'} |n'\rangle |n\rangle\langle n'|
 \label{matrixel}
\end{equation}
acts on the space of the multi-electron nanographene states, with $N=N_s$.
The matrix elements are given by:
\begin{equation}
 \langle n|\hat{{\cal O}}^{(+)}_{\alpha\alpha'}|n'\rangle = \sum_{ii'm_+}\frac{V_{\eta\sigma}(i)V^*_{\eta'\sigma'}(i')}{E_{m_+}-E_0-\bar{\epsilon}^{\eta\eta'}_{nn'}}\gamma^{m_+}_{nn'}(ii'\sigma\sigma')
\label{Oplusel}
\end{equation}
and
\begin{equation}
 \langle n|\hat{{\cal O}}^{(-)}_{\alpha\alpha'}|n'\rangle = \sum_{ii'm_-}\frac{V^*_{\eta\sigma}(i)V_{\eta'\sigma'}(i')}{E_{m_-}-E_0+\bar{\epsilon}^{\eta\eta'}_{nn'}}\gamma^{m_-}_{nn'}(ii'\sigma\sigma')
\label{Ominusel}
\end{equation}
where $\bar{\epsilon}^{\eta\eta'}_{nn'}=\frac{\mu_\eta+\mu_{\eta'}+\Delta_{nn'}}{2}$, $\Delta_{nn'}=E_n-E_{n'}$ and $\gamma^{m_\pm}_{nn'}$ are  the spectral function weights, given by equation (\ref{gamma}).

\begin{figure*}
 \centering
  \includegraphics[width=0.90\textwidth]{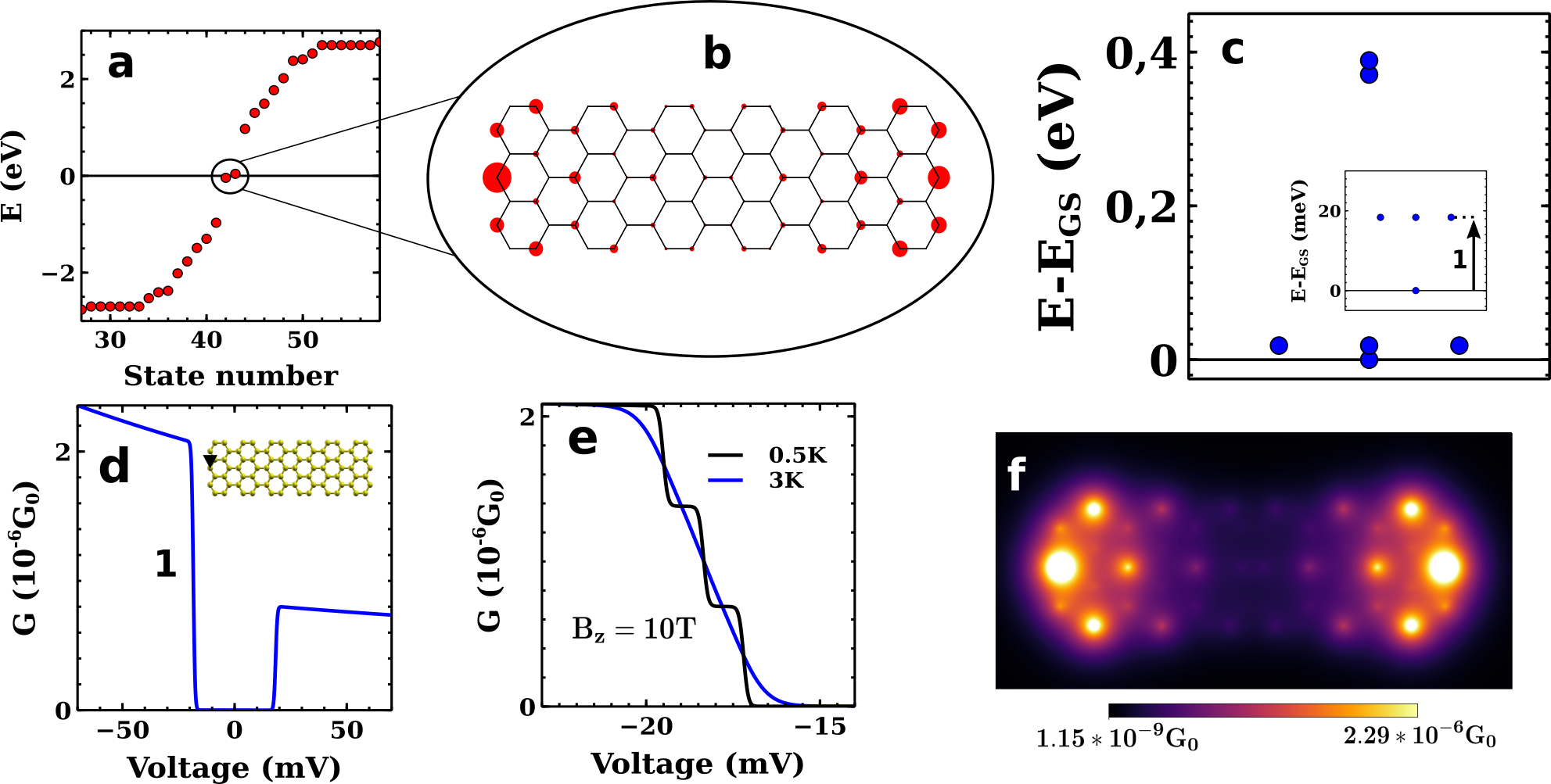}
 \caption{ Tight-binding  single-particle spectrum (a) and  square of the wavefunction for the lowest energy in-gap molecular orbital for the nanoribbon with $t =-2.7eV$,  and $\delta t=-0.3 t$ for the atom
 at the left edge marked in the inset of panel (d). Notice that, because of the tip-induced
 deformation, the molecular orbital has  different weights at left and right edges. 
c) The first 6 eigenstates calculated with the Hubbard model for $U = |t|$ and CAS(2,2). The inset is a zoom, showing a transition from the 
ground state to the first excited triplet state, labeled with 1. 
d) dI/dV curve. The singlet-triplet transition is seen as a step,  
the black triangle points the atom where the tip is coupled.
$V_S = t/100$ and $V_T = V_S/10$,
 $T=3K$, $\rho_{\eta\sigma}=10/t$ and 
$\epsilon_C = -5.7eV$.
e) Conductance curve for two different temperatures and an off-plane magnetic field ($B_z$). By applying a magnetic field, the Zeeman effect splits step 1 in three. f) Map of the nanoribbon conductance when $V_{bias}=-40mV$. }
\label{ribboncot}
\end{figure*}

In the case of cotunneling, we assume   $P_n$ is given by the Boltzmann equilibrium functions, and we ignore their voltage dependence thereby.  Finally, the calculated current  (equation \ref{intens}) is given by the scattering rates ($W^{\eta\eta'}_{nn'}$) between different neutral states labeled as $n$ and $n'$, which depend on
 temperature, bias,  electrode density of states ($\rho_{\eta\sigma}$), central system-electrode coupling, and central system wavefunctions. The formula reads as
\begin{equation}
 I_{T\rightarrow S} = e \sum_{nn'} P_n (W^{S\rightarrow T}_{nn'} - W^{T\rightarrow S}_{nn'})
 \label{intens}
\end{equation}
where the scattering rates are then given by:
\begin{equation}
 W^{\eta\eta'}_{nn'}\approx \sum_{\sigma\sigma'}\frac{2\pi\rho_{\eta\sigma}\rho_{\eta'\sigma'}}{\hbar}{\cal G}(\mu_\eta - \mu_{\eta'}+\Delta_{nn'})\Sigma^{\eta\sigma,\eta'\sigma'}_{nn'}
 \label{W}
\end{equation}
where ${\cal G}(x) = \frac{x}{1-e^{-\beta x}}$, $\beta = \frac{1}{k_BT}$, and:
\begin{equation}
\Sigma^{\eta\sigma,\eta'\sigma'}_{nn'}= |\langle n|\hat{\cal O}^{(+)}_{\eta\sigma,\eta'\sigma'}-\hat{\cal O}^{(+)}_{\eta'\sigma',\eta\sigma}|n'\rangle|^2
\label{Sigma}
\end{equation}

So, to sum up, the calculation of the cotunneling conductance is carried out through the following steps:
\begin{enumerate}
\item Solution of the single-particle model to find the molecular orbitals of a given nanographene.
\item Solution of the many-body problem, in a restricted space of configurations defined by states with integer occupation of the molecular orbitals, in the manifolds with $N=N_s, N_s\pm 1$. 
\item Calculation of the matrix elements in equation (\ref{gamma}), the  effective cotunneling Hamiltonian elements in
eqs.(\ref{Oplusel},\ref{Ominusel},\ref{Sigma}), the scattering rates in equation (\ref{W}), that permit to compute the cotunneling  current (\ref{intens}) for a given bias.
\end{enumerate}

\section{Cotunneling spectroscopy of spin excitations in nanographene diradicals}
We now discuss  the cotunneling inelastic electron tunneling spectroscopy (IETS) of three  representative nanographenes:
a rectangular nanoribbon\citep{wang2016}, triangulene\citep{pavlivcek2017} and Clar's goblet\citep{mishra19b} (see figure \ref{fig2}a).
The three of them are diradicals\citep{Ortiz18,Ortiz19}.
 Clar's Goblet and the ribbon have both a  $S=0$  ground state whilst the triangulene
has a $S=1$ triplet. Here we will infer in how this technique results to be useful to demonstrate this spin quantum number, and in last instance
the sign of the exchange for the expected local moments for these molecules.

\subsection{Rectangular graphene nanoribbons}

We first consider a rectangular graphene nanoribbon ([6]ribbon in table\ref{table1}) as those  reported by Wang et al.\citep{wang2016} This system
presents two in-gap quasi-zero modes\citep{Ortiz18} inside a large gap (figure \ref{ribboncot}a), 
whose wavefunction is strongly localized at the zigzag edges (figure \ref{ribboncot}b). 
This system provides an effective realization of a Hubbard dimer\citep{Ortiz18}, 
governed by two energy scales: the hybridization energy, measured by the splitting of the in-gap states $\delta$, and the effective Hubbard repulsion, given by\citep{Ortiz19}:
\begin{equation}
{\tilde U}= U \sum_i |A(i)|^4
\end{equation}
where $A(i)$ is the amplitude of the in-gap edge mode at site $i$ in the $A$ sublattice. The effective Hubbard repulsion is relatively independent of the ribbon width. In contrast, the hybridization energy depends exponentially.  In the limit $\tilde{U}>> \delta$,   the ground state is a correlated spin singlet, separated from an excited triplet state by\citep{Anderson59,Ortiz19} 
\begin{equation}
J\simeq \frac{\delta^2}{\tilde{U}}
\end{equation}
Our CAS calculations corroborate this picture (figure \ref{ribboncot}c).  For $U=|t|$ we obtain $J=  18 meV$. The dependence of $J$ on $U$ is shown in figure (\ref{stepsvsU}a) of the Appendix\ref{hubbardposition}.

The cotunneling conductance, shown in figure (\ref{ribboncot}d),   features an inelastic step at $|e|V_{bias}=\pm J$, when the tip is placed in an atom where the edge modes have sufficiently large amplitude (panels (b) and (f) in figure \ref{ribboncot}).  This is ascribed to tunneling events in which energy is given from the tunneling quasiparticles to excite the singlet-triplet transition of the antiferromagnetically correlated edge states.

The singlet to triplet nature of the inelastic step can be  confirmed   upon application of a magnetic field that splits the triplet \citep{Hirji06}.   As a result, the  inelastic step splits according to the rule 
\begin{equation}
\Delta=J+ g\mu_B B_z S_z
\end{equation}
where $S_z$ can take 3 values,  $S_z=-1,0,+1$.
This leads to the appearance of 3 steps, instead of only 1, in the cotunneling conductance shown in figure (\ref{ribboncot}e).

The emergence of  three steps unveils the $S=1$ nature of the excited state. 
In the case  of a triplet  ground state, relevant for the triangulene discussed below, additional steps appears instead at low bias $|e|V_{bias}= \pm g\mu_B |B_z|$, coming from inelastic excitations within the ground state manifold. Therefore,  the absence of this feature, along with the splitting of the inelastic step and the selection rules, is enough to determine the degeneracies of the ground state and first excited state of the graphene nanoribbon.

Scanning the inelastic step intensity across the nanographene provides an  additional tool to explore the collective spin excitations of the nanographene with the STM. As we show in figure (\ref{ribboncot}f),  the height of the conductance inelastic step  highly depends on the lateral spatial position of the tip.
In figure (\ref{ribboncot}f)  we show  the map for $V_{bias}=-40mV$ with $B_z=0$. The IETS scan has  a direct correspondence with the wavefunction of the quasi-zero modes, further confirming  the  edge nature of the collective excitations. 

In the discussion section we comment on why the inelastic steps predicted here have not yet been observed experimentally in rectangular nanographenes \citep{wang2016}.

\subsection{Clar's goblet}
\begin{figure}[h!]
 \centering
  \includegraphics[width=0.5\textwidth]{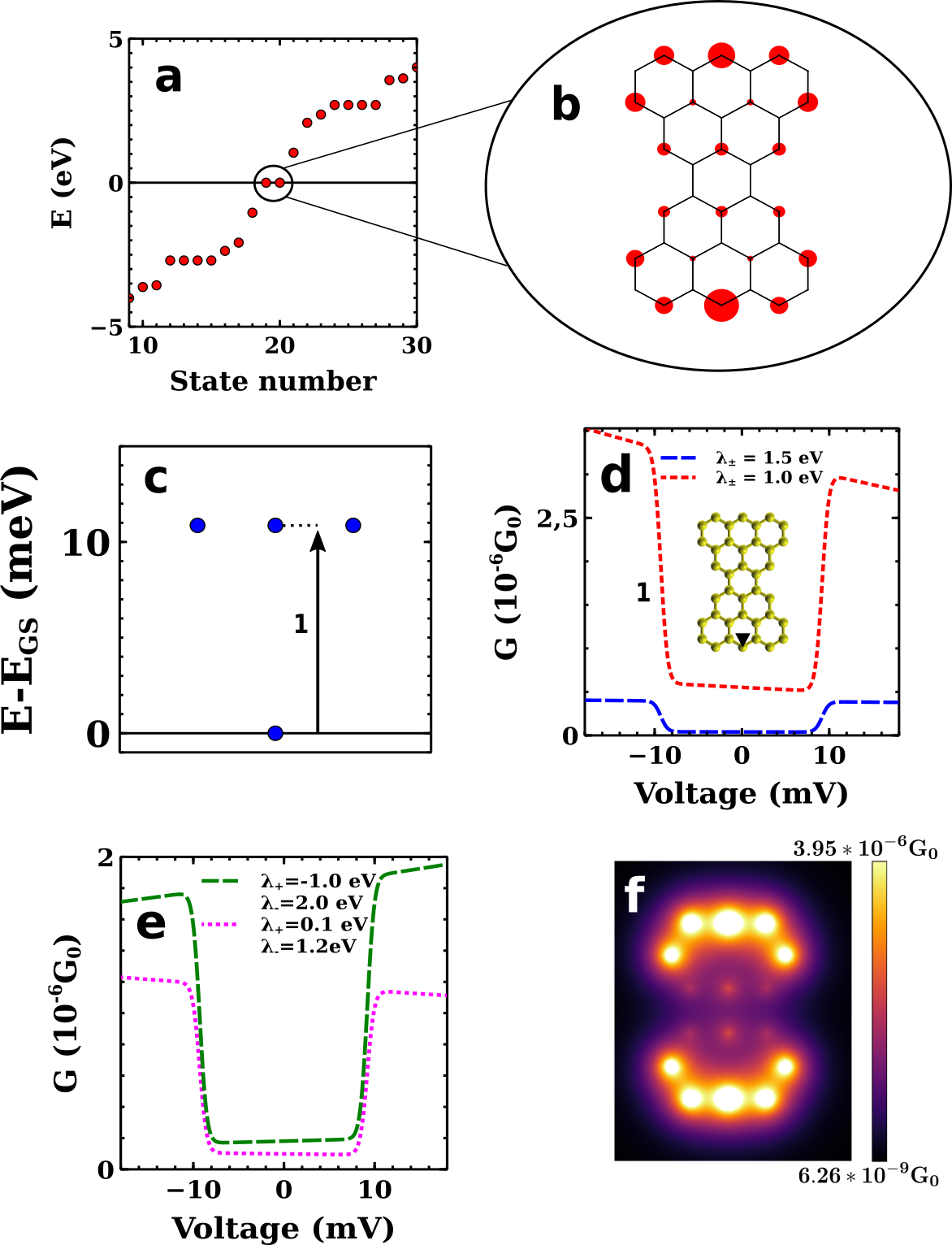}
\caption{a) Tight-binding single-particle spectrum for the Clar's goblet with $t =-2.7eV$ and $\delta t=-0.3 t$ for the atom
 at the bottom edge, marked in the inset of panel (d), and b) is the corresponding wavefunction for the zero modes. 
c) The ground state and first excited triplet calculated with the Hubbard model for $U = |t|$ and CAS(4,4). The transition from the former to
the latter is represented by the arrow and labeled by 1. 
d) dI/dV curve with
$\lambda_\pm = 1.0eV$ (red) and $\lambda_\pm = 1.5eV$ (blue). The black triangle points the atom where the tip is coupled. e) dI/dV curve for different dominant channels. 
$V_S=t/100$, $V_T=V_S/5$, $T=3K$, $\epsilon_C = -5.7eV$, $\rho_{\eta\sigma}=10/t$ and $\mu_S=-5.4eV$.
f) Map of the Clar's goblet conductance when $V_{bias}=-15mV$.}
\label{bowcot}
\end{figure}

We now apply cotunneling  theory to the case of Clar's goblet or graphene bowtie.
Clar's Goblet\citep{clar72}, shown in figure (\ref{fig2}a),  is a diradical nanographene  with $S=0$ ground state, as expected\citep{Ortiz19} on account of its lack of sublattice imbalance. The single-particle spectra features both bonding and anti-bonding states, separated by a large gap,  plus two in-gap   zero modes that arise
from the  fusion of two graphene fragments, with one zero mode each, that remain unhybridized   when they are fused to form the bowtie (figure \ref{bowcot}a,b).

The many-body spectra
 is very similar to the case of the rectangular graphene nanoribbon discussed in the previous section (see figure \ref{bowcot}c). However,  the singlet-triplet splitting does not arise from kinetic exchange, as $\delta=0$ for the bowtie. Exchange arises here from correlations that involve the virtual excitations of  single-particle states other than the in-gap zero modes, and scales with $U^2$\citep{Ortiz19}.
 
 For the model  considered here, we obtain a  singlet-triplet splitting $\Delta=11 meV$, for $U=|t|$. If we include up to third neighbour hopping\citep{Ortiz19}, then $\Delta$ increases. 
  The dependence of $\Delta$ on $U$ and on the third neighbour hopping $t_3$ is shown in
   figure (\ref{stepsvsU}b) of
    Appendix\ref{hubbardposition}.
  First principles methods\citep{Ortiz19} give $\Delta= 20meV$, whilst experiments   give $23meV$. 
  However,  all these calculations ignore the coupling to the substrate that might remormalize the exchange in nanographenes.

Our results for cotunneling spectroscopy are shown in figure (\ref{bowcot}d) for $\lambda_{\pm}=1.0$ and $\lambda_{\pm}= 1.5eV$.  It is apparent that, as $\lambda_\pm$ is increased, the height of the inelastic step decreases, as expected given that cotunneling amplitude scales inversely with the addition energies. In contrast,  the energy of the step does not depend on $\lambda_{\pm}$,  as $\lambda_{\pm}$ affect the addition energies but not the excitation energies of the $N=N_s$ manifold.
We also took $\mu_S$ as the work function of the electrode, that was chosen as -5.4 eV for Au. 

 In figure (\ref{bowcot}e) we explore the dependence of the dI/dV as we change $\lambda_{\pm}$ to make the addition or removal manifolds lower in energy, deciding thereby the virtual channel, either electron or hole, that  controls cotunneling. We find that the slope of the cotunneling curve changes, depending on the nature of the dominant cotunneling channel. This  originates from the bias dependence of the addition and substraction energies, that controls the cotunneling amplitudes (see equations \ref{Oplusel},\ref{Ominusel}).

In figure (\ref{bowcot}f) we map the intensity of the inelastic steps across the bowtie structure and we find it matches the wavefunction of the zero modes, very much like in the case of rectangular nanographenes.

\begin{figure*}
 \centering
  \includegraphics[width=0.80\textwidth]{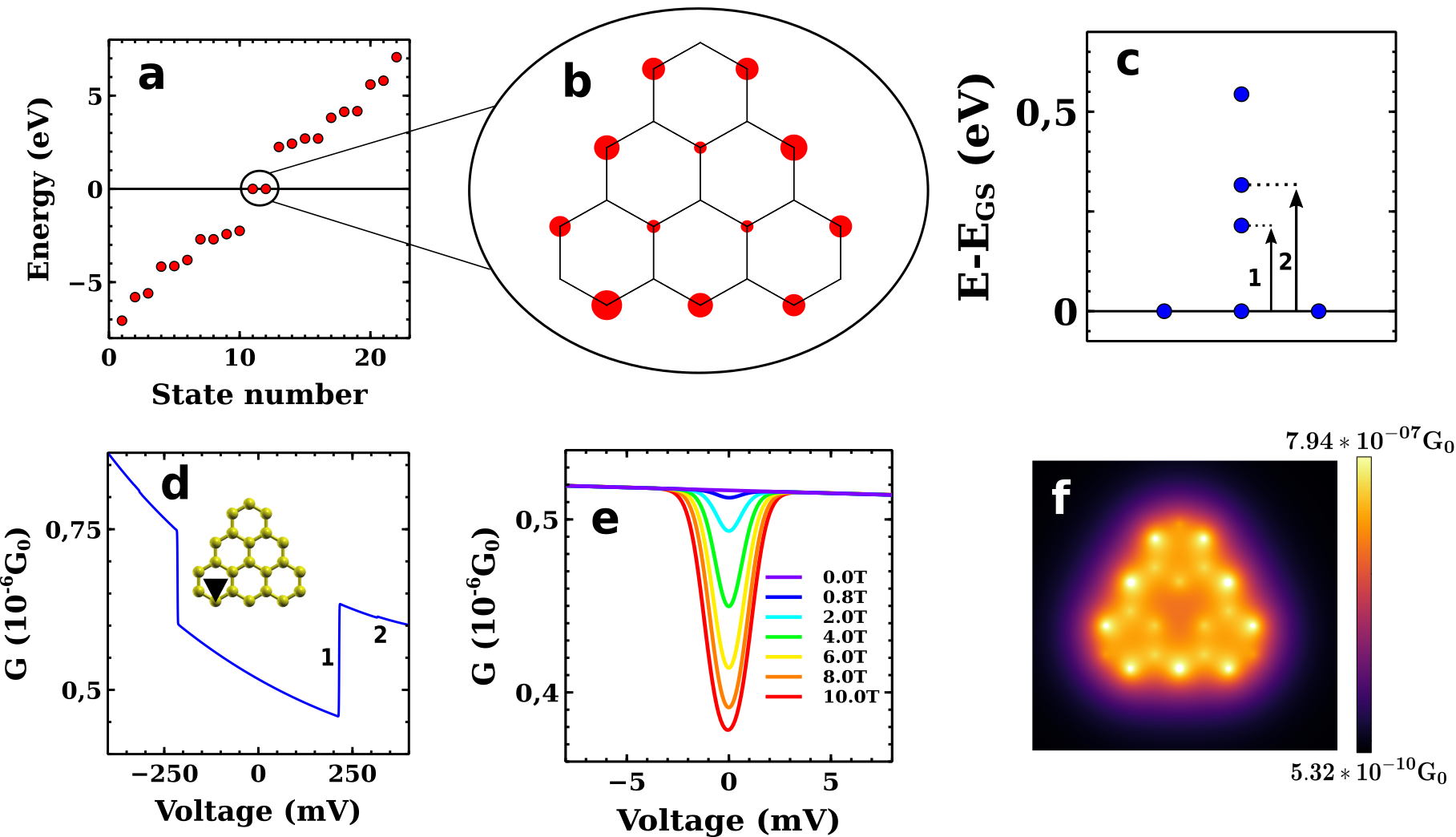}
 \caption{a) Tight-binding single-particle spectrum for  triangulene with $t =-2.7eV$ and $\delta t=-0.3 t$ for the atom
 at the bottom edge marked in the inset of panel (d), and b) the corresponding wavefunction for the zero modes. 
c) The first 6 eigenstates calculated with the Hubbard model for $U = |t|$ and CAS(4,4). Transitions from the ground state to the 2 first singlet
excited states are represented as arrows and labeled by 1 and 2. 
d) dI/dV curve. The two first transitions from the ground state to the singlet excited states are seen as two separated steps,  
the black triangle points the atom where the tip is coupled. $V_S=t/50$, $V_T=V_S/10$, $T=3K$, $\rho_{\eta\sigma}=10/t$ and 
$\epsilon_C = -5.7eV$.
e) Conductance curve with different off-plane applied magnetic fields, a dip around 0 bias appears as a consequence of the Zeeman effect. f) Map of the triangulene conductance when $V_{bias}=-310mV$.}
\label{triangcot}

\end{figure*}

\subsection{Triangulene}

We now consider the triangulene shown in figure (\ref{fig2}a) with 22 carbon sites. 
This is known to be a diradical with $S=1$\citep{inoue01}.
%
Unlike the bowtie and the rectangular nanographene,  the triangulene considered here  has a sublattice imbalance, with 10 atoms in one sublattice and 12 in the other. As a result, the single-particle spectra features two sublattice polarized in-gap zero modes\citep{Palacios08,Lado15,Ortiz19}, as shown in figure (\ref{triangcot}a).  Their wavefunctions (figure \ref{triangcot}b) can be chosen as eigenstates of the $C_3$ symmetry operator\citep{Ortiz19}. The molecular orbitals of these $C_3$ symmetric zero modes   have the same modulus\citep{Ortiz19}, and therefore a maximal overlap, that enhances ferromagnetic exchange.

The sublattice imbalance $N_A-N_B=2$  implies\citep{Lieb89, JFR07} that the ground state of the triangulene, described with the Hubbard model at half filling and first neighbour hopping, has  $S=1$. The many-body spectra, calculated with the CAS(4,4) approximation has a $S=1$ 
ground state followed by two degenerate singlets and then one more singlet.
The peculiar degeneracy of the lowest energy excitation is a consequence of the $C_3$ symmetry. 

When we compute the conductance for this symmetric configuration, we obtain an extremely small height of the inelastic step at the energy of the lowest triplet-singlet excitation. 
We can gain some insight on the origin of this result in the case of CAS(2,2).  In this case, it  can be  seen  that the contribution to the 
cotunneling matrix elements  of the coupling between the triangulene and the substrate is proportional to $\sum_i\phi_z(i)$. Interestingly, the  zero modes of undistorted triangulene can be chosen to  satisfy the identity $\sum_i\phi_z(i)=0$, which automatically gives a vanishing cotunneling conductance.  A finite, but small, conductance is obtained in the CAS(4,4) approximation.

The cotunneling conductance is further increased if 
we break the $C_3$ symmetry, by assuming that the atom underneath the tip is pulled out of the surface, reducing its hopping with the first neighbours in the triangulene (see equation \ref{def}).  Upon this approximation the resulting single-particle wavefunction (figure \ref{triangcot}b) has a larger weight on the atom underneath the tip, and is no longer true that $\sum_i\phi_z(i)=0$.

The resulting many-body spectra for the distorted triangulene  is shown in figure(\ref{triangcot}c). 
As a result of the distorsion, the excited states with $S=0$ are no longer forming a doublet, as obtained in the  $C_3$ symmetric case\citep{Ortiz19}.
The cotunneling conductance of the deformed triangulene, with $B_z=0T$, is shown in 
figure (\ref{triangcot}d).  It has two steps, corresponding to the inelastic excitation of the triangulene  $S=1$ ground state towards the deformation-split excited doublet with $S=0$.

Application of a magnetic field brings an effect specific of the $S=1$ ground state.
Because of the Zeeman splitting, 
  the state with magnetic moment parallel to the applied field becomes predominantly dominated at low temperature, and the others become depleted.  This entails a reduction of the elastic cotunneling contribution that diminishes  the zero bias conductance. In addition, a new finite bias inelastic step appears for $|e|V_{bias}=\pm g\mu_B |B_z|$. These two effects are  shown in figure (\ref{triangcot}e).    Observation of this feature would provide a conclusive confirmation of the $S>0$ nature of the ground state. We believe this effect has been observed by Li  and coworkers\citep{li2019}, although they also observe a $S=1$ Kondo peak that can only be captured if we treat interaction with the substrate going beyond the second order perturbation theory discussed above\citep{jacob16}.

In figure (\ref{triangcot}f) we show 
 the map for the dI/dV signal 
 for $V_{bias}=-310mV$.  As in the case of bowtie and rectangular nanoribbons, mapping the inelastic conductance provides an additional variable to probe the excitations and relate them to the zero modes, whose molecular orbitals are peaked at the  edges.

\section{Discussion}

\subsection{Conditions for the observation of inelastic cotunneling steps}
Cotunneling  theory predicts the observation of cotunneling steps in every structure for which the inelastic excitation of the $N=N_s$  manifold is significantly smaller than the addition and substraction energy peaks.   Whereas this type of excitations have now been reported in several structures\citep{NachoNat,li2019,mishra19b} they are conspicuously  missing in all triangulenes  \citep{pavlivcek2017,su19,mishra19} as well as in the rectangular ribbons\citep{KimoucheNat2015,wang2016,Ruffieux16}.

There are several  factors that are known to reduce the visibility of cotunneling steps:
\begin{enumerate}
\item Thermal broadening\citep{lambe68}, approximately given by $5.4 k_B T$.  This could blur the inelastic steps  of low energy excitations, such as the Zeeman split ground to ground transitions in the triangulene, or  the singlet to triplet transitions of long rectangular ribbons for which exchange energy  decays exponentially with size \citep{Ortiz18}.
\item Lock-in voltage.  The resolution of the inelastic steps cannot be better than the lock-in voltage. Therefore, application of lock-in voltages larger than the inelastic excitations compromises their visibility. 
\item Excitation lifetime effects.  Inelastic tunneling spectroscopy is probing the  spectral function of the collective excitations\citep{JFR09}. The poles of this spectral function are broadened by the inverse of their lifetimes. Kondo coupling to the substrate results in a broadening proportional to the energy of the excitation\citep{delgado10,khajetoorians11}.
\item Competition with sequential  processes. The sequential tunneling is a resonant process, and therefore contributes much more strongly to the  conductance. This could shadow inelastic steps whose energy is not sufficiently different from the sequential peak. 
\end{enumerate}

In order to address the lack of experimental observation of cotunneling steps in  $S=1$, $S=3/2$ and $S=2$  triangulenes,
that have been recently synthetized and probed with STM\citep{su19,mishra19}, 
we have computed their  triplet-singlet splitting. For  $U=|t|$ the excitation energies are $\Delta=260,192,146meV$ for $S=1$, $S=3/2$ and $S=2$, respectively.
In figure (\ref{stepsvsU}c,d) of the Appendix\ref{hubbardposition} we show the linear dependence of these energies on $U$.
  Inspection of the experimentally observed values for the $V_-$ peaks show that the visibility of the negative bias step for $S=3/2$ might be compromised.

\subsection{Symmetry and bias dependence}
The cotunneling theory naturally yields curves that, in addition to the step-like features when the bias matches the excitation energies of the system,   can have both a superimposed finite voltage dependence  away from the inelastic steps and/or  a very different height of the steps for positive and negative bias.  We refer to these two features as bias dependence and asymmetry. 

Bias dependence is a consequence of the fact that the matrix elements in equations (\ref{Oplusel},\ref{Ominusel}), that determine
the effective tip-surface tunneling amplitudes in the cotunneling Hamiltonian(\ref{cotun}), feature  the {\em electrode-averaged}  addition energies in the denominators\citep{DelgadoPRB2011}.     When the bias voltage is comparable with some of the addition energies, these denominators could cancel. The cotunneling theory is only valid far from these values. However,  even when the bias is away from  these addition energies, these denominators bring a voltage dependence that  only disappears  when bias voltage is sufficiently far.

The cause of the asymmetry of the height of the steps at positive and negative bias is quite different. The ultimate origin of the asymmetry is the fact that coupling of the nanographene to the tip and the substrate is not symmetric.  Hand wavingly,    the process by which an electron tunnels from tip to one carbon atom and then tunnels from any atom of the nanographene towards the substrate has a very different amplitude from the reverse process, by which an electron tunnels from surface to any atom in the nanographene and then it tunnels from one carbon atom to the tip.  

Numerical evidence of these statements is presented in the  
figure (\ref{stepasym}) in Appendix\ref{assym}.  Figure (\ref{stepasym}a) shows a symmetric $dI/dV$ cotunneling
curve for a Hubbard trimer  symmetrically coupled to both electrodes. Figure (\ref{stepasym}b) shows a mildly asymmetric conductance when the two electrodes are connected to only one atom in a non-equivalent configuration.  Finally,  an extremely asymmetric conductance is obtained when the coupling to the electrodes is very different (see figure \ref{stepasym}c).

\subsection{Kondo}
As discussed above, the observation of zero bias peaks at low temperature in some nanographenes
\citep{NachoNat, Ezawa2009, li2019, mishra2020} provides a very strong evidence of the emergence of local moments in graphene and their quenching via Kondo interaction with the substrate. The  Hamiltonian  (\ref{cotun}) is a generalized Anderson model that can in principle describe Kondo correlations if we compute current to higher order in the graphene-electrode interaction\citep{jacob16}.  This will be the subject of future work.

\section{Summary and outlook}

We have revisited both  sequential tunneling  and cotunneling theories, often used to model transport in  quantum dots\citep{LossPRB2004} and molecules\citep{gaudenzi17} , and discussed their application to 
study STM spectroscopy of nanographenes on surfaces. The main goal is to understand how the STM  $dI/dV$ measurements can convey information  about the existence of local moments in the $\pi$ electrons of nanographenes.   We have discussed in detail the case of  three classes of  diradical nanographenes that have been recently studied experimentally: rectangular nanoribbons, bowtie and triangulenes.   

An important take home message is that sequential transport leads to peak features in $dI/dV$ that relate to the addition and substraction energies of the nanographenes and provide thereby no direct information of the spin states. Sequential transport can be used to infer the addition energy of a given frontier orbital, a quantity that plays a role in the formation of local moments in nanographenes. 

 In contrast,  cotunneling spectroscopy provides direct information of the energies of  neutral  excitation  of nanographenes.  These energies can provide direct evidence of the formation of local moments, if supplemented with magnetic field dependence experiments.  Specifically, cotunneling spectroscopy could be in principle used 
to determine if  the spin of the ground state is finite: for   $S>0$ ground state, application of a magnetic field can result in a dip at zero bias. 
STM  also permits us to   map the inelastic signal  as the STM tip scans the molecules laterally.  
As discussed above, the $dI/dV$ maps so obtained  have a similar profile than 
zero modes that host the unpaired electrons that form the local moments, highlighting their interconnection.

Both theory and STM experiments probing open shell nanographenes indicate that emergent moments are governed by large energy scales.   In the case of triangulenes, the $S=1$ ground state is separated from $S=0$ excitations by a gap of several hundreds of meV, although this has not been observed experimentally yet. 
The $S=0$ ground state of  Clar's goblet is separated by the $S=1$ excited state by    23 meV, which implies an inter-molecule exchange interaction of that magnitude\citep{Ortiz19}. 

Advances in on-surface synthesis make it possible to assemble larger structures that combine open shell fragments with finite spin and  large intermolecular exchange interactions and to explore strong coupling carbon based structures,  with exchange interactions that can be both ferro and antiferromagnetic. Given both the small magnetic anisotropy of carbon based magnetism\citep{Lado2014prl},  as well as their low dimensionality (0D, 1D or at most 2D),  this kind of artificial structures will  provide an ideal platform to explore quantum magnetism, very much like the case of  magnetic adatoms\citep{choi19,khajetoorians19}. 

Future theory work should address a more realistic treatment of the nanographene-substrate interaction that is able to provide a theory of the linewidth of both sequential and cotunneling  features,  the observation of Kondo effect\citep{NachoNat,li2019},  the effect of substrate induced spin relaxation and decoherence\citep{delgado17}, and the renormalization of the  excitation energies due to Kondo coupling to the substrate\citep{oberg14}.  The interplay between local moments in nanographenes and proximity induced superconductivity\citep{san15,Lado2DShiba} should also be a fertile arena to discover exotic new phases of matter.

{\em Acknowledgments}
We acknowledge  A. P\'erez-Guardiola, G. Catarina, J. C. Sancho-Garc\'ia, M. Shantanu,  O. Gr\"oning,  P. Ruffieux,  R. Fasel,   and J. I. Pascual for fruitful discussions.
We acknowledge financial support from MINECO-Spain (Grant No. MAT2016-78625-C2) and  from  the Portuguese Funda\c{c}\~ao para a Ci\^encia e a Tecnologia  (FCT) for the projects P2020-PTDC/FIS-NAN/4662/2014,
 P2020-PTDC/FIS-NAN/3668/2014 and
 UTAPEXPL/NTec/0046/2017 projects. JFR acknowledges  Generalitat
Valenciana funding (Prometeo2017/139).
 R. O. acknowledge ACIF/2018/175 (Generalitat Valenciana and Fondo Social Europeo).

\appendix

\section{Dependence of the excitation energies  on $U$}
\label{hubbardposition}

\begin{figure}[h!]
 \centering
  \includegraphics[width=0.45\textwidth]{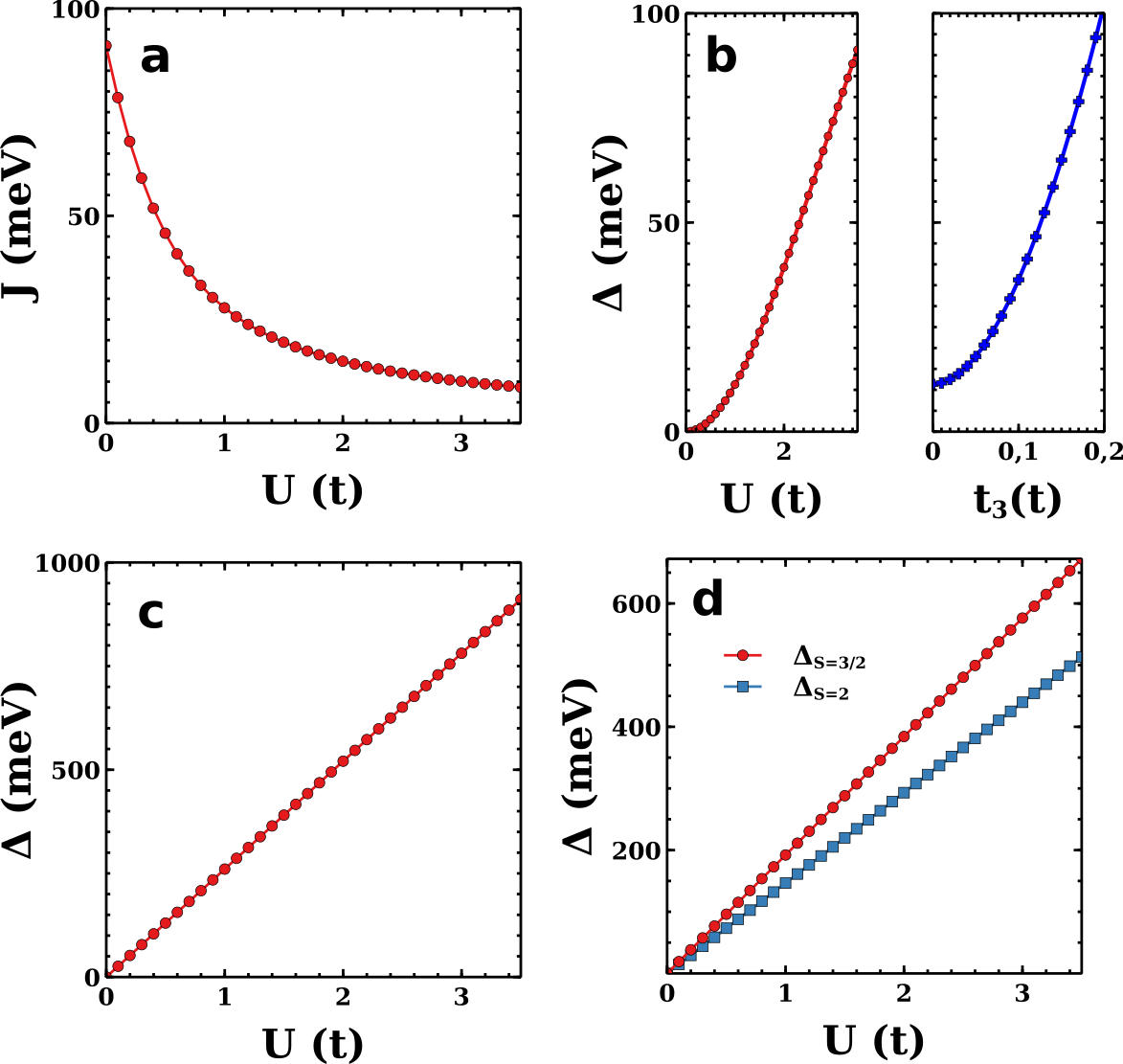}
\caption{Splitting between the ground state and the first excited state for a) nanoribbon (CAS(2,2)), b) Clar's goblet (CAS(4,4)), c) $S=1$ triangulene (CAS(2,2)) and d) $S=3/2$ (CAS(3,3)) and $S=2$ (CAS(4,4)) triangulenes, as function of $U$. 
$t = -2.7 eV$ and $\epsilon_C = -5.7eV$. No hopping distortion was introduced here.
For the Clar's goblet, the dependence with third neighbour hopping is also displayed in panel right with constant $t_2=-0.071eV$ and $U=|t|$.}
\label{stepsvsU}
\end{figure}

In this Appendix we show the dependence on $U$ of the excitation energies  in the $N=N_s$ manifold for the 3 classes of systems considered in the paper (nanoribbon, bowtie, triangulene). These enegies determine  the  bias voltage at which  inelastic steps  appear.  The results are shown in  figure (\ref{stepsvsU}). The different $U$ dependence is discussed in the main text. In the case of the bowtie we also compute the dependence of the excitation energy on the third neighbour hopping $t_3$ (see  right panel of figure \ref{stepsvsU}b).

\section{Asymmetric inelastic steps height
\label{assym}}

\begin{figure}[h!]
 \centering
  \includegraphics[width=0.45\textwidth]{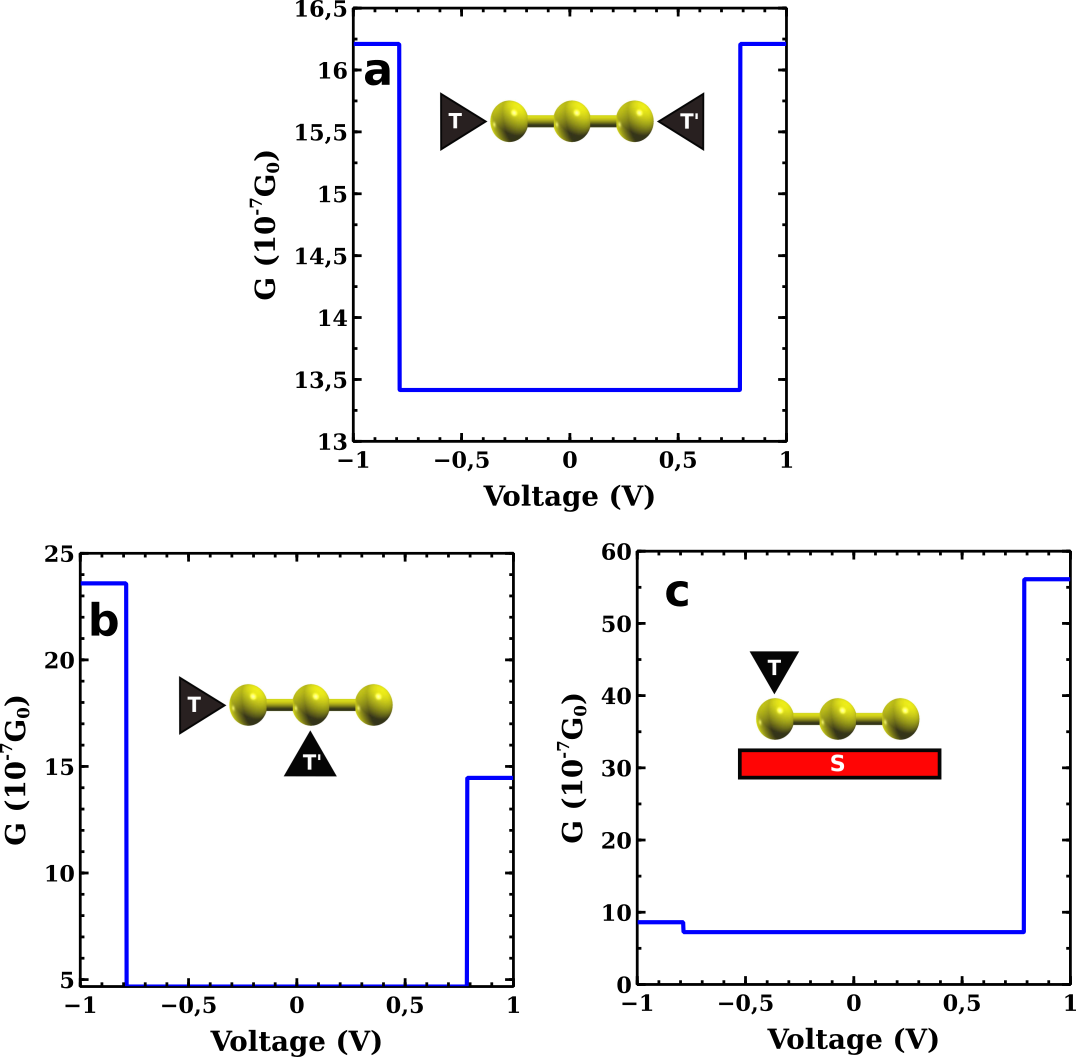}
\caption{dI/dV curves for the Hubbard trimer when a) two tips are linked to the side atoms, b) one tip is linked to a side atom and a second tip to
the center atom, and c) one electrode is a tip and the other is a surface. CAS(3,3), $E_F=0eV$, $t=-1eV$, $U=4|t|$, $Ed=-2.0eV$, $V_{S,T'}=t/50$, $V_T=V_S/10$, $\rho_{\eta\sigma}=10/t$ and $T=1K$. Here
both $\mu_{\eta}$ and $\mu_{\eta'}$ are a function of bias, so the bias dependence is avoided for clarity. This calculation was performed without hopping distortion.}
\label{stepasym}
\end{figure}

As discussed in the main text,  cotunneling spectra can feature different height in the inelastic steps for positive and negative bias.  This is clearly seen in the rectangular ribbon, and less clear in the triangulene. The qualitative reason is discussed in the text and is ultimately due to the asymmetry of the
coupling of the nanographene to tip and surface . A numerical validation of this statement is shown in 
figure (\ref{stepasym}).

\newpage

\bibliographystyle{naturemag}
\bibliography{biblio}{}
\end{document}